\documentclass[aps,preprintnumbers,amsmath,amssymb,floatfix,groupedaddress,nofootinbib]{revtex4}

\usepackage{enumerate}
\usepackage{graphicx}
\usepackage{color}
\usepackage[utf8]{inputenc}

\usepackage{fontenc}  
\usepackage{amsmath}

\usepackage[francais]{babel}
\newcommand{\beq}{\begin{equation}}
\newcommand{\eeq}{\end{equation}}
\newcommand{\bea}{\begin{eqnarray}}
\newcommand{\eea}{\end{eqnarray}}
\newcommand{\ba}{\begin{array}}
\newcommand{\ea}{\end{array}}

\newcommand{\bef}{\begin{figure}}
\newcommand{\eef}{\end{figure}}

\begin{document}

\title{Completing  the quantum formalism in a contextually objective framework. }

\author{Philippe Grangier}

\affiliation{ \vskip 2mm
Laboratoire Charles Fabry, IOGS, CNRS, Universit\'e Paris~Saclay, F91127 Palaiseau, France.}

\begin{abstract}
In standard quantum mechanics (QM), a state vector $| \psi \rangle$ may belong to infinitely many different orthogonal bases, as soon as the dimension $N$ of the Hilbert space is at least three. On the other hand, a complete physical observable $A$ (with no degeneracy left) is associated with a $N$-dimensional orthogonal basis of eigenvectors. In an idealized case, measuring $A$ again and again will give repeatedly the same result, with the same eigenvalue.   Let us call this repeatable result a modality $\mu$, and the corresponding eigenstate $| \psi \rangle$. A question is then: does $| \psi \rangle$ give a complete description of $\mu$~? 

The answer is obviously no, since $| \psi \rangle$ does not specify the full observable $A$ that allowed us to obtain $\mu$; hence the physical description given by $| \psi \rangle$ is incomplete, as claimed by Einstein, Podolsky and Rosen in their famous article in 1935.  Here we want to spell out this provocative statement, and in particular to answer the questions: if $| \psi \rangle$ is an incomplete description of $\mu$, what does it describe~? is it possible to obtain a complete description, maybe algebraic ?  Our conclusion is that the incompleteness of standard QM is due to its attempt to describe systems without contexts\footnote{ The present article is self-contained, but it elaborates on the publications \cite{csm4a,csm4b,csm5} where more background can be found.}, whereas both are always required, even if they can be separated outside the measurement periods.
\end{abstract}

\maketitle


\section{Introduction.  } 

In previous articles \cite{CO2002,csm1,csmBell,csm2,csm3,csm4a,csm4b,csm5}   
we introduced an approach to quantum mechanics (QM) where the relevant physical objects are not ``quantum systems", but ``quantum systems within classical contexts". These extended objects carry physical properties that are certain and repeatable, and that are  called modalities; hence the acronym CSM for Contexts, Systems and Modalities.  
Let us illustrate these definitions with a small quizz, starting with the two statements : 
\vskip 2mm

(a) Alice  prepares  a spin 1/2 particle in the pure state $| + \rangle_{\vec u}$ , eigenstate of $\vec S. \vec u$; then  she sends the particle to Bob.  Without knowing $\vec u$, Bob cannot identify the state  $| + \rangle_{\vec u}$ without error, though this state  is known for sure by Alice (she can repeatedly measure it as long as she owns the particle). Therefore the state appears as Alice's subjective knowledge. 

(b) Alice  prepares a spin 1/2 particle within a context,  in a modality corresponding to the result $+\hbar/2$ for a measurement of  $\vec S. \vec u$; then she sends to Bob the particle {\bf and} the context, classically defined by the direction $\vec u$. Now Bob can find $+\hbar/2$  with certainty by measuring $\vec S. \vec u$. The modality is thus an objective property of the particle within a context. 

Now the quizz: should we speak about the subjective state of the system as in (a), or about the objective modality of a system within a context as in (b) ? 
\vskip 2mm

Usual QM gives the first answer, whereas CSM gives the second one. The CSM answer fits with the idea  that {\bf a physical property is objective (i.e. belongs to an object) if anybody who is given the object can determine the value of this property from a measurement}. In classical physics, an isolated system is an object, and it owns measurable properties. In quantum physics, objective properties belong to a system within a context, because both of them are needed to make the property objective \cite{CO2002}. This is surprising at first sight : after the preparation, the system may be fully isolated from the context, still the context is required to define the modality, as illustrated in the quizz above, where getting the system alone does not allow Bob to identify its ``state". In practice, as well as in principle, it is not required that Bob gets Alice's full Stern-Gerlach apparatus, because this apparatus can be classically described, and the only relevant adjustable parameter is the value of $\vec u$.
\vskip 2mm

It is important to note that for an Hilbert space of dimension $N=2$, like in the quizz above, there is only one context admitting  $|+ \rangle_{\vec u}$ as a basis vector, together with $|- \rangle_{\vec u}$; therefore the value of $\vec u$ is enough to specify the context.  But if  $ N \geq 3$ there are infinitely many contexts that include a given $|\psi \rangle$, because any set of $N$ orthonormal vectors including $|\psi \rangle$ defines a context.  As a simple example, consider  two spins 1/2 with operators $\vec S_1$ and $\vec S_2$: the eigenstate $|++\rangle$ for $(S_{1z}, S_{2z})$ is the same ``state" as  the eigenstate $|S=1, m=1\rangle$ for $(\vec S^2, Sz)$, where $\vec S = \vec S_1 + \vec S_2$; however the two contexts are incompatible; therefore $|++\rangle$ and $|S=1, m=1\rangle$ correspond to different modalities. 
\vskip 2mm

Looking now at the issue from the other side, what is a ``state" $|\psi \rangle$, such as $|++\rangle \equiv |S=1, m=1\rangle$,  with respect to a modality ? It should be clear that $|\psi \rangle$ is a mathematical object associated with an ensemble of modalities belonging to different contexts, in such a way that these modalities are connected with certainty when the context is changed. This defines an equivalence relation between modalities, that we call extravalence; generically, this connexion between modalities is called extracontextuality; this plays an essential role to get Born's rule from Gleason's theorem \cite{csm4b}.
\vskip 5mm

These simple observations strongly suggest that the usual quantum state, when defined as a vector $|\psi \rangle$ an Hilbert space, is incomplete: though it does allow to predict the result of all future measurements on the system, it is not ``recoverable" given the system alone - and thus it is not objective according to our previous definition. But what is missing ? Certainly not some mysterious hidden variables, but simply the context itself : in the usual formalism, the state vector should be completed by the full measurement set-up, or by the complete set of commuting observables (CSCO) admitting $|\psi \rangle$ as an eigenstate. Actually, as already pointed out by Bohr \cite{EPR,Bohr1935}, this is done implicitly everywhere in standard QM; here we argue that doing it explicitly would remove many wrong ideas. 
\vskip 2mm

To conclude this introduction,  CSM does not contradict usual QM, but leads to quite different views about its meaning, such as the  idea of  completing the quantum formalism by combining 
classical and quantum observables. This is in opposition with the idea that classical physics  should ``emerge'' from quantum physics, and as well in opposition with the idea that there should be some ``hidden variables'', lying below the quantum observables, and giving more details about them.  
In the next sections we shall proceed in to build up a complete mathematical formalism, including both systems and contexts; and then we'll draw some consequences about standard QM.

\section{Contexts in quantum mechanics: from implicit to explicit.  } 
\vskip - 2mm

\subsubsection{Completing the quantum formalism: why to do it.}
\vskip - 2mm

We will start from the idea of an ``isolated system'', that  is a sub-entity of the natural world, that can be given some definite physical properties : position, mass, velocity, angular momentum...  for particle-like subsystems; but it may also carry amplitude, phase, frequency... for wave-like subsystems. The goal of physics is to identify such systems, to define their properties from both the physical and mathematical points of view, and to predict their evolution and interactions. It is taken for granted at the theoretical level that these systems are isolated well enough to carry such properties, but it is important to note - though it is not usually emphasized -  that they are immersed into a broader physical world, allowing measurements to be carried out, results to be registered, etc.  
\vskip 2mm

{\bf  A basic difference between classical and quantum physics  is how an isolated system should be described and managed}. In QM the properties of the system can be perfectly well defined, i.e. they can be made reproducible and predictable with certainty, but this requires to include also some relevant properties of the broader physical world around the system, that we will generically call the {\bf context}.  In order to distinguish such a reproducible set of properties from the usual ``quantum state of the system", we will call it a {\bf modality}.  
\vskip 2mm

It is then clear that 
two different types of properties are available at the context level to specify a modality: those specifying a macroscopic property of the context (e.g. the orientation $\vec u$ of the Stern-Gerlach magnet), and those specifying the result obtained within this context (e.g. $+$ or $- \hbar$/2). The different values ($\vec u$, $\pm$) are well defined at the context level, no ``super context'' is needed to define them better:  the broader physical world is the same in classical and quantum physics. 
Nevertheless, significant differences appear between the classical and quantum situations :

- in classical physics, an isolated system ``owns'' its physical properties, and the context can be ignored: though it is certainly there, its role remains  accessory. 

- in quantum physics the modalities ($\vec u$, +) and ($\vec v$, +) are obviously different when  $\vec u$ and $\vec v$ are different, but $\vec u$ and $\vec v$ are context's (and not system's) observables. In addition, $\vec u$ and $\vec v$ cannot be realized simultaneously within a given context, this would be counterfactual.  It is not satisfactory either to say that the system carries the value of $\vec u$ as a hidden variable, because $\vec u$ cannot be recovered given the system alone (and there are many more impossibility theorems, including Bell's \cite{Bell,Aspect,Laloe}). So the classical and quantum formalisms are different - but can we unify them ?
\vspace{-4mm}
\subsubsection{The usual QM formalism and the CSM approach. }
\vskip - 2mm

In  the usual QM formalism, the observables are operators within a non-commutative algebra $\{ A_m \}$, where $A_m$ is a $N \times N$ matrix if there are $N$ mutually exclusive modalities in any given context. The indices $\{ m \}$ can be seen as a generic notation for all the context parameters specifying the operator $A_m$. Without loss of generality, one can assume that $A_m$ is non degenerate, i.e. equivalent to a complete set of commuting observables (CSCO). One gets thus a set of rank-1 orthogonal projectors $\{ Q_{m, i = 1...N} \}$ on the eigenstates of $A_m$, and the projectors $\{ Q_{m,i} \}$ specify a quantum state $| \psi_{m,i} \rangle$, associated with some eigenvalue $a_{m,i}$ of $A_m$. From the spectral theorem, $A_m = \sum_i a_{m,i} \; Q_{m,i}$. 
\vskip 2mm

The CSM approach does not start directly from this formalism, but obtains it by induction, from two basic physical rules, that have been introduced and discussed in details in previous articles \cite{CO2002,csm1,csmBell,csm2,csm3,csm4a,csm4b}:

\noindent - {\bf contextual quantization:} for a given system within a given context, the maximum number $N$ of mutually exclusive results (defined as modalities)  is a property of the system, and it is the same in any relevant context. This value of $N$ is also the dimension of the matrices $\{ A_m \}$ or $\{ Q_{m,i} \}$ introduced above, which gets now a clear meaning. 

\noindent - {\bf extracontextuality of modalities:} modalities belonging to different contexts may be connected with certainty, and this transfer of certainty  defines an equivalence class between modalities, called extravalence \cite{csm3}.  This extravalence, obvious from empirical evidence, is a crucial piece of the  construction, because it defines a connection between contexts, which appears when $N \geq 3$.  It also pinpoints an essential property of QM that is not taken properly into account, neither by the usual formalism, nor by the usual terminology on contextuality \cite{context}. 
\vskip 2mm

Then it is demonstrated in \cite{csm4b} that 
given an initial modality and context, obtaining another modality in another context follows a probabilistic law, 
with a probability which keeps the same value for given initial and final extravalence classes.
This result suggests to associate a mathematical object to an extravalence class, and the projectors $\{ Q_{m,i} \}$ perfectly fit that need ! Then it appears that all the hypotheses of Gleason's theorem are fulfilled, so Born's rule follows \cite{csm4b}.
It has been shown in other articles \cite{csm2,csm3} that complex numbers must be used, and that all representations (or bases) within the resulting $N$ dimensional Hilbert space are unitarily equivalent; this is just the probabilistic framework of usual QM, with $N$ finite or countably infinite. 
Describing the system using $N \times N$ matrices acting on a Hilbert space appears thus as a mathematical construction, induced from the empirical observations of quantization in $N$ mutually exclusive modalities in each context, and extravalence between modalities from different contexts.  
\vskip 1mm

Though this picture appears quite satisfactory, it would be better to describe the systems and contexts in a unified mathematical framework. It is clear however that neither the usual quantum nor the classical formalisms can be used, because the first one fails to describe classical contexts, whereas the second one fails to describe quantum systems. It is therefore required to extend the mathematical tools to be used. 
\vspace{-4mm}
\subsubsection{Some basic notions about operator algebras. }
\vskip - 2mm

The basic intuition for the required extension comes from an article published in 1939 by von Neumann, and entitled ``Infinite Direct Products" \cite{Jvon Neumann1939}. Direct products are now called tensor products, and they are generally used in QM for combining the Hilbert spaces associated with different systems (and contexts). The idea used by von Neumann is  to study mathematically the limit where the number of such systems becomes infinite. Naively, one would expect to get an ``infinitely large Hilbert space", still with the same algebraic properties, but this turns out to be completely wrong. Quoting von Neumann: {\it 
Infinite direct products differ essentially from the finite ones in this, that they split up into ``incomplete direct products". (…) An essential result of our theory is that the ring of all the bounded operators which are generated (algebraically or by limiting-processes) by operators of the individual algebra does not contain all bounded operators acting on the infinite tensor product.  (…) What happens could be described in the quantum-mechanical terminology as a splitting up of the tensor product into ``non-intercombining systems of states", corresponding to the incomplete direct products quoted above."}
Later these ``incomplete direct products" have been called superselection sectors, and these ideas have been rewritten in an even more mathematical form, called operators algebra. But the main idea does remain: the limit corresponding to a (countably) infinite tensor product of separable Hilbert spaces is uncountably infinite, and not separable. Therefore it is not manageable in practice, and it ``splits up" in sectors, that can be defined by commuting macroscopic observables: this means they behave classically. 
\vskip 2mm

Using now a modern algebraic langage, we need a non-classical probability theory based on a non-commutative measure theory, and it is well known \cite{aqm} that non-commutative von Neumann algebras can be associated with non-commutative measure spaces. In addition, any von Neumann algebra is generated as a norm closed subspace by the set of the spectral projections corresponding to its self-adjoint elements. We will therefore consider von Neumann algebras rather than $C^*$ algebras (any von Neumann algebra  is a $C^*$ algebra, but not the reverse), and focus on projectors and unitary operators, rather that on the usual self-adjoint observables (observables can be obtained from eigenvalues and  projectors through the spectral theorem).  If we associate projectors with events in our non-classical probability theory, the question is then: what are the available projectors  in a von Neumann algebra ? 
\vskip 2mm

Again, any von Neumann algebra $\cal M$ is generated as a closed subspace by the set $P(\cal M)$ of all orthogonal projections in $\cal M$, and $P(\cal M)$ contains all spectral projections of any normal operator contained in $\cal M$. A first step into the structure of von Neumann algebras is therefore the analysis of its set of projections, and a few definitions are  in order. 
First, a factor is a von Neumann algebra $\cal M$ that has a trivial center, i.e. $Z(\cal M) = {\cal M'} \cap {\cal M} = \mathbb{C} \;  {\cal I}$ (the commutant $\cal M'$ of $\cal M$ is the set of all bounded linear operators on H commuting with every operator in $\cal M$, and $\cal I$ the identity). 

So let $\cal M$ be a  factor, then we have the following classification (due to Murray and von Neumann \cite{aqm}):
\vskip 2mm

\begin{itemize}
\item $\cal M$ is of type I if it contains a minimal, non-zero projection (a projection $P$ is minimal if $Q \leq P$ implies $Q = 0$ or $Q = P$ ; one has $P \leq Q$ if the image of $P$ is contained in the image of $Q$, or equivalently if $PQ = P$). 

In this case we say that $\cal M$ is of type I$_n$, $n \in N \cup \{\infty\}$, if ${\cal I} = \sum_{\ell =1}^n P_\ell$, where $\{P_\ell\}_{\ell =1}^n$ is a family of minimal projections in $P(\cal M)$. If $n = \infty$, then the infinite sum is understood in the strong operator  topology. Standard QM is generically using type I$_n$ factors, with  I$_\infty$ corresponding e.g. to algebras acting on wave functions. 
\vskip 2mm

\item $\cal M$ is of type II if it contains no minimal projections, but has finite non-zero projections. In this case we say
that $\cal M$ is of type II$_1$ if $\cal I$ is finite (i.e. $\cal M$ is finite), and that  $\cal M$ is of type II$_\infty$ if $\cal I$ is infinite (i.e. $\cal M$ is infinite). Type II$_1$ was considered by von Neumann as the natural extension of type I, but is now seldom used in physics. 
\vskip 2mm

\item $\cal M$ is of type III if it contains no finite non-zero projection. The finer classification of type III factors into factors of type III$_0$, of type III$_\lambda, \; \lambda \in (0,1)$, or of type III$_1$, uses deep results in Modular Theory and we don’t enter in this issue here.  Type III is used in quantum field theory and statistical physics, for systems with an infinite number of particles or degrees of freedom. 
\vskip 2mm

\end{itemize}

Factors of type I are completely classified up to isomorphism: any factor $\cal M$ of type I is isomorphic with the set $L({\cal H})$ of bounded operators on some Hilbert space $\cal H$, so the dimension $N$ of $\cal H$ is a complete invariant for factors of type~I. Again, this is the familiar situation in usual QM, fitting with the $N$ mutually exclusive modalities of CSM.
\vskip 2mm

Factors  of type II (with no minimal projection) or  type III (with no  finite non-zero projection) are not intuitive from a physical point of view. We will not enter in more details here, but an important issue is that the usual tensor product remains well defined in von Neumann algebras. Therefore it makes sense to consider the mathematical object $Q \otimes P$, where $Q$ belongs to a usual type I algebra, whereas $P$ belongs to a type III algebra; then the tensor product is also in a a type III algebra. As we will see below such objects are essential in our way to complete QM. 
\vspace{-4mm}
\subsubsection{Completing the quantum formalism: how to do it.}
\vskip - 2mm

Within the algebraic formalism, let us introduce an extended  {\bf commutative} algebra $\{ Q_{m,i} \otimes P_{m,i} \}$, where $P_{m,i}$ is an (infinite dimensional) projector on the state of the context associated with a result $i$ within the context parameters $m$, which are defined at the classical level. All the operators $\{ Q_{m,i} \otimes P_{m,i} \}$ commute, and correspond to a classical level of description where there are $N$ mutually exclusive modalities  in each context.  
\vskip 2mm

Using the algebraic terminology, and in the absence of any measurement, each value of $(m,i)$ corresponds to a (continuous) superselection sector. An essential feature is that such a sector can be identified by only a few classical parameters, embedded in the values of $(m,i)$, though a quantity like a polarizer's orientation involves  a macroscopic object  with myriads of atoms moving in a cohesive way, and coupled to various quantum fields. The corresponding algebra does not fit in the usual type I in  the above classification \cite{emch}, because the projectors $P_{m,i}$ are operating in a space with non-countably infinite dimension.  Then, as written above, superselection sectors appear as a generic mathematical property \cite{AQFT,OA}, imposed by the algebraic formalism  
describing the relevant context's physics. It is a quite noticeable feature that, despite the high complexity of this algebra, only a few macroscopic classical observables are enough to specify the relevant data, i.e. to distinguish one sector $P_{m,i}$  from another \cite{ng}. 
\vskip 2mm

The extended  algebra $\{ Q_{m,i} \otimes P_{m,i} \}$  allows us to be ``factual" (as opposed to counterfactual) at the context level, since all observables commute and are classically defined. In each context a measurement gives one among $N$ mutually exclusive results,  
and only one result $i$ is realized at a time in context $m$,  i.e. that counterfactuality is not allowed.  On the other hand, due to extracontextuality, the same projector $Q_{m,i}$ may appear in different contexts; said otherwise, there may be contexts descriptions $P_{n,j}$ such that  $Q_{m,i} = Q_{n,j}$, so extravalence naturally fits in the picture. 
\vskip 2mm

Admitting now that the extended commutative algebra $\{ Q_{m,i} \otimes P_{m,i} \}$ is the fundamental mathematical object to completely describe a system within a context,  the next  question is~: since we need to  recover the usual practice of QM, is it possible to ``forget" about the $P_{m,i}$, and bring back all the $Q_{m,i}$ in the same $N \times N$ Hilbert space~? 
\vspace{-3mm}
\subsubsection{Going back into the Hilbert space.}
\vskip - 2mm

For doing so we have to consider quantum measurements, that have been ignored so far. The previous construction corresponds to uncoupled modalities, and it can be said it is a kinematic rather than dynamical structure. Nevertheless, Born's rule can be obtained from Gleason's theorem, showing clearly that the probability rule comes only from the geometry of the (projective) description of probabilities, without requiring any dynamical elements. It can be said that 
changing modalities has ``no duration", but in physical terms this duration should be seen as infinite rather than zero,  because the modalities are asymptotic, i.e. they are defined ``long before" or ``long after" the measurements. 
\vskip 2mm

Then the question arises: how to describe the measurement itself, i.e. the change from some modality  $Q_{m,i} \otimes P_{m,i}$ to another one $Q_{n,j} \otimes P_{n,j}$ ? Clearly the dynamics should involve additional algebraic elements or projectors, but they cannot be built by ``superposition" of the available ones, due to the sectorization. Generally speaking, time evolution within a type III algebra is very difficult to manage mathematically, and this is somehow vexing since the asymptotic states are already known, as well as the probabilistic Born's rule connecting them when the context is changed. 
\vskip 2mm

The physical solution to this problem is well known in practice:  the measurement can be described in an approximate way, by considering  the apparatus as a large collection of ``ancillas" coupled with the initial system, and getting entangled with it. It is essential to remember that such a large ``pseudo-measuring device" is always embedded in a context, where the system's asymptotic modalities are already identified, from  the very definition of the quantum system's $N$-dimensional Hilbert space. Therefore such a modality must be recovered at the end of the process, if it has any meaning as a quantum measurement on the isolated system of interest. 
\vskip 2mm

But then, as long as the measurement does not ``really happen" as a proper quantum jump (i.e., reaching a new asymptotic modality), the coupling with more and more ancillas may be described as a unitary evolution of the initial modality $Q_{m,i} \otimes P_{m,i}$, where there system and context evolve essentially independently, respecting the algebraic sectorisation.  More precisely, the evolution of $Q_{m,i}$ interacting with ancillas will obey Heisenberg-type equations, and the evolution of $P_{m,i}$ will follow the classical evolution at the context level.  
\vskip 5mm

These evolutions are in principle reversible, including entanglement and dis-entanglement with ancillas, and the effect of external Hamiltonian like a classical magnetic field acting on a spin. When the measurement really happens, associated with irreversibility \cite{csm4a}, the sectorisation temporarily blows up, but it is known in advance that the system and context will end up in another modality, as prescribed by Born's rule. 
In summary: 

- the future behaviour of the system depend only on the projectors $Q_{m,i} $, whereas the $P_{m,i}$ are needed for making the mathematical description of the modality objective and complete;

- the extended commutative algebra $\{ Q_{m,i}  \otimes P_{m,i} \}$ is not type I, because the projectors $P_{m,i}$ are operating in a space with non-countable infinite dimension. Crucially, this space is ``sectorized", as it is usual in  non-type I algebras \cite{AQFT}. Physically this corresponds to the fact that  a unique ``context-and-result" is defined outside measurement periods;

-  on the other hand, the (non-commuting) observables $A_m = \sum_j a_{m,i}  \, Q_{m,i} $  for the system are acting in a separable Hilbert  space with either finite or countably infinite dimension (type~I), and the projectors $Q_{m,i} $ correspond to extravalence classes of modalities. Formally this is just usual QM, though it is dressed in a quite different way. 
\vspace{-4mm}
\subsubsection{Looking again at the ``cut" between system and context.}
\vskip - 2mm

From the above it also becomes clear that the so-called ``cut" between the system ($Q_{m,i}$) and the context ($P_{m,i}$) is simply a consequence of the very definition of an isolated system.  As mentioned before, a quantum isolated system differs fundamentally from a classical one, because its physical properties (modalities) are quantized in a given context; therefore it is impossible to ignore this context, as it would be the case in classical physics. 
The cut is moveable depending on the definition of the system, but it must  fulfill well-defined conditions: 
\begin{enumerate}
\item there is only one cut, because by construction  the $P_{m,i}$ are the only projectors acting in an unbounded space. It is however possible to increase the size of the system and to redefine its dimension $N$,  moving the cut accordingly;
\item in order to get a consistent unitary evolution of the projectors $Q_{m,i}$, the choice for the cut,  i.e. the  definition of the $Q_{m,i}$'s  and $P_{m,i}$'s, must be agreed upon a priori,  following the definition of the system;
\item there is no restriction on the size of the system, but it must always be embedded in a context, in agreement with the two previous rules. Correspondingly, there is always a context ``at the infinity edge".
\end{enumerate}

Let us emphasize again that we impose the commutation of the global observables $\{ Q_{m,i} \otimes P_{m,i} \}$ on the  basis of physical arguments, and obtain as a result the quantum description of isolated systems within contexts, {\bf outside} of measurements periods. Since the observables $\{ Q_{m,i} \otimes P_{m,i} \}$ are of no practical use during a measurement, one can simply use the standard quantum formalism, building a dynamical model of the system-context interactions, and making suitable approximations. Then it can be shown \cite{balian} that ``the pointer observables behave as if they commuted with any observable of the algebra; however, this holds only in the final states (i.e., when the measurement is fully completed)". This shows the consistency between  the present approach, starting from general physical principles and implementing them within the algebraic formalism, and a more dynamical approach, starting from usual QM and making suitable approximations. In both cases,  there is only one ``factual state" of the macroscopic world, and the ``full quantum state" (or modality) of a subsystem  belongs jointly to a quantum system and  a classical context.

\section{ Caveats and benefits. } 
\vskip - 2mm

\subsubsection{Compatibilities and incompatibilities with other approaches.}
\vskip - 2mm
To summarize, our claim is that the combination of (i)~the physical (ontological) framework given by contexts, systems and modalities (CSM), and (ii)~the mathematical framework given by the algebraic quantum formalism, provides a consistent account of how QM is currently working. Obviously not everybody is using CSM or C$^*$ algebras, so does our claim make any sense ? 

On the CSM side, the proposed ontological reference to a ``system-within-a-context"  is quite compatible with  usual textbook quantum mechanics, but  it clashes with some other interpretations, endowed with either strong ontologies (e.g.  Everett-DeWitt many worlds, or de Broglie-Bohm), or no ontology at all (e.g. qbism).  

On the algebraic side, AQFT has been criticized for not being able to recover perturbative results based on renormalization, obtained within the standard model. The situation is still evolving \cite{AQFT,BF}, but here we are more concerned in low-energy QM, and with issues related with the measurement problem or quantum non-locality. Here the algebraic framework works quite satisfactorily, and it eliminates unjustified extrapolations of textbook QM, that were already rejected by Von Neumann \cite{Jvon Neumann1939}, but are unfortunately still flying around. It can be said  that textbook quantum mechanics is not universal, because it is using a quite restricted mathematical framework; but by extending it in a quite reasonable (algebraic) way, many of these erroneous extrapolations simply disappear. 
\vspace{-4mm}
\subsubsection{Removal of the ``infinite regression'' .}
\vskip - 3mm
For instance, there is no more ``infinite regression'' leading to generalized macroscopic entanglement, because mathematically the infinite limit does not produce a pure vector state, but a statistical mixture (i.e. the relevant representation is reducible).  Let us emphasize that the issue here is not to have a {\it physically infinite} number of particles, but to assert that to describe appropriately the limit of a macroscopically large number of particles in a measurement context, it is required to be consistent with a {\it mathematically infinite} number of particles. This is why an ``infinite regression" obtained from the usual type I formalism is simply meaningless. As a reminder, all the above statements make sense within a probabilistic interpretation of the QM formalism~: the physical reality corresponds to the contexts, systems and modalities, and the quantum formalism as described above is a method to calculate probabilities, taking into account the rules of quantization and extracontextuality. 
\vspace{-4mm}
\subsubsection{Removal of the ``ambiguity of mixtures''.}
\vskip - 3mm
Our approach removes also the  ``ambiguity in the decomposition of a mixed state into pure states" \cite{balian}. If a vector state  is defined in the global space, it will automatically involve statistical mixtures of modalities corresponding to different sectors, unless the representation is irreducible and the state is pure. But there is no ambiguity in such vector states (pure or mixed), since all relevant context parameters are specified in each sector; irrelevant parameters do not matter by construction.  So the ambiguity appears only when this statistical distribution is imported back in the $N \times N$ Hilbert space, removing thus the context observables  (see III.2). From this restricted space there are many ways to reconstruct a mixture  for non-pure states, it is even possible to reconstruct pure states by adding an entangled ancilla, as it is done in standard purification procedures. This is one more evidence that limiting the description to a $N \times N$ Hilbert space is a convenient mathematical construction, but not  a complete description of the real modalities. 
\vspace{-4mm}
\subsubsection{Removal of the ``preferred basis problem''.}
\vskip - 3mm
Similarly, the question of the ``preferred basis choice" in a quantum measurement is readily solved by considering that most of the context parameters (e.g. the polarizer orientation...) cannot be modified by the interaction Hamiltonian describing the measurement.  Actually, given an initial modality, it is convenient to consider that the operators $\{ A_m \}$  evolve deterministically and unitarily in the Heisenberg picture, either due to internal interactions, or due to external deterministic control; this corresponds to what is required for a ``quantum computer" evolution. If this evolution is well controlled, and as long as no new measurement is made, it will always be possible to determine  a context where the result of the measurement can be predicted with certainty: this is what the Heisenberg evolution is telling us, while the modality remains the initial one.   This smooth evolution stops if a new measurement is performed in a context with incompatible modalities, then Born's rule and sectorization apply, and the process is re-initialized in the new context with a new set of macroscopic observables $\{ P_{m,i} \}$. 

Summarizing, a measurement requires that the quantum  system is no more isolated, but coupled to the external world; this means  one has to jump out from the initial modality, and to get a new one once the measurement is completed in a new context. The overall ``jump" probability is given by Born's law, and Gleason's theorem shows that it has a geometrical origin, given the empirically imposed constraints of quantization and extracontextuality~\cite{csm4b}. The projection postulate is a convenient recipe to summarize this complicated process, but it has no ontological implication, i.e. there is no physical ``reduction of the wave packet" - but rather quantum jumps between modalities. 
\vspace{-4mm}
\subsubsection{Reconsideration of Bell's inequalities. }
\vskip - 3mm
Another noteworthy remark is that by completing QM in this more general framework, Bell's inequalities cannot be written any more, because they involve a counterfactual mix up of different classical contexts. In a more explicit way, let us consider a standard Bell-CHSH situation  \cite{Aspect}, with two entangled particles $a$ and $b$, and measurements of  $A_1$ or $A_2$ for $a$, $B_1$ or $B_2$ for $b$. One has $[A_m, B_j] = 0$ for $i,j=1,2$, and in order to get a violation of Bell's inequalities {\bf both} commutators $[A_1, A_2] $ and $[B_1, B_2]$  must be nonzero \cite{baez}. In a more general way, both algebras $\{ A_m \}$  and $\{ B_j \}$ must be non-commutative. 

But if one consider the extended algebras for $a$ and $b$, 
which are commutative and sectorized, there is no more violation in this extended algebra.  
In some sense,  the violation of Bell-CHSH inequalities is never ``measured'', but inferred assuming that the statistics are determined by the properties of the system (the entangled pair) in a context-free way. This idea comes essentially from classical physics, where the system owns context-free properties; 
this is why  only local and {\bf classical} realism is tested by Bell's inequalities. Obviously the contextual objectivity \cite{CO2002} (or quantum realism) used by CSM \cite{csm1} leads to a completely different point of view, not constrained by Bell's inequalities. 
\vspace{-4mm}
\subsubsection{Other issues and conclusion.}
\vskip - 3mm

Generally speaking, physical theories consist essentially of two elements, a kinematical structure describing the states and observables of the system, and a dynamical rule describing the change of these states and observables with time. In the present article we focused on  the kinematical part, including Born's law, but it is also needed to specify the dynamical law governing the change with time of the observables or states. This allows a system-specific description of the measurements, including e.g. the calculation of decoherence times and similar quantities. More work may be done within the algebraic approach, but it is already known from various other methods \cite{aqm,balian} that the idealized asymptotic states, outside the measurement periods, do match the kinematical pure states considered here. 

Does this approach change anything to the way QM is currently used ? Not much as far as practical questions are concerned, 
but  recurrent foundational  issues like  the ``measurement problem'' or ``spooky influences at a distance'' are simply removed. The algebraic formalism is also suitable for connecting usual QM to quantum field theory \cite{emch,AQFT,BF}. In more general settings involving gravity, there is no problem in considering that the space-time metric is defined at the classical (context) level, as it is the case in standard general relativity.  
\vskip 3mm

Any hint about quantum gravity ? A possible way to go beyond classical  general relativity would be to look for a ``suitably isolated quantum gravitational system'', but it is not clear whether such a notion may even be defined, because now the ``isolated system'' would have to include the space-time metric itself. 
Quantization and extracontextuality make sense for any standard measurable  physical quantity, but it is unclear also how they might be applied to the metric itself. Our very tentative conclusion is that full (non-linear) quantum gravity matters only at Planck's scale, where all physical interactions have little to do with our low-energy experience. 
\vskip 3mm

So is it required to learn about CSM and operator algebras~? The answer is that it depends on your interests. In practice, it is quite possible to do quantum physics without worrying about the measurement problem or spooky actions at a distance; but in case you do worry, have a look at the above, it may help you with such concerns. Then it should be emphasized that both physics and mathematics are needed, with quite separate roles: either mixing them up, or considering one without the other,  may very likely bring you back to some well-known dead ends. 
\vskip 2mm

{ \bf Acknowledgements.}
The author thanks Alexia Auff\`eves,  Nayla Farouki, Franck Lalo\"e and Roger Balian for many useful discussions.


\begin{references}

\bibitem{CO2002} P. Grangier, ``Contextual objectivity: a realistic interpretation of quantum mechanics'', European Journal of Physics 23:3, 331 (2002) [arXiv:quant-ph/0012122]. 

\bibitem{csm1} A. Auff\`eves and P. Grangier, ``Contexts, Systems and Modalities: a new ontology for quantum mechanics'', Found. Phys. 46, 121 (2016) [arXiv:1409.2120].


\bibitem{csmBell}  A. Auff\`eves and P. Grangier,  ``Violation of Bell's inequalities in a quantum realistic framework'', Int. J. Quantum Inform. 14, 1640002 (2016) [arXiv:1601.03966].

\bibitem{csm2} A. Auff\`eves and P. Grangier,  ``Recovering the quantum formalism from physically realist axioms'', Scientific Reports 7, 43365 (2017) [arXiv:1610.06164].

\bibitem{csm3}  A. Auff\`eves and P. Grangier,  ``Extracontextuality and extravalence in quantum mechanics",  Phil. Trans. R. Soc. A 376, 20170311 (2018) [arXiv:1801.01398].

\bibitem{csm4a}  A. Auff\`eves and P. Grangier, ``A generic model for quantum measurements", 
Entropy 21, 904 (2019), https://www.mdpi.com/1099-4300/21/9/904, [arXiv:1907.11261].  

\bibitem{csm4b}  A. Auff\`eves and P. Grangier, ``Deriving Born's rule from an Inference to the Best Explanation", Found. Phys. (2020),  https://doi.org/10.1007/s10701-020-00326-8, [arXiv:1910.13738]. 

\bibitem{csm5}  P. Grangier, ``Completing the quantum formalism: why and how ?",  https://arxiv.org/abs/2003.03121v2

\bibitem{EPR} A. Einstein, B. Podolsky, and N. Rosen, ``Can Quantum-Mechanical Description of Physical Reality Be Considered Complete?'', Phys. Rev. 47, 777 (1935).

\bibitem{Bohr1935} N. Bohr, 
same title as \cite{EPR}, Phys. Rev. 48, 696 (1935).

\bibitem{Bell}  J.S. Bell,  ``On the Einstein-Podolski-Rosen paradox'', Physics 1, 195 (1964).

\bibitem{Aspect}  A. Aspect,  ''Closing the Door on Einstein and Bohr's Quantum Debate'', Physics 8, 123 (2015)

\bibitem{Laloe}  F. Lalo\"e, ``Do We Really Understand Quantum Mechanics?'', Cambridge University Press (2012).

\bibitem{context} In QM the assignment of {\bf values} to measurement results is contextual (Kochen-Specker theorem), whereas the assignment of {\bf probabilities} to measurement results is non-contextual (Gleason theorem, see also \cite{csm4b}). This creates endless confusions in the terminology, and extracontextuality is a good way out of them. For projectors eigenvalues 0 and 1 are contextual, whereas average values (probabilities) are non-contextual; this is another way to tell that there is no dispersion-free state. 

\bibitem{Jvon Neumann1939} J. von Neumann, ``On infinite direct products",  Compositio Mathematica 6, 1-77 (1939).

\bibitem{ng}  In \cite{Jvon Neumann1939} an incomplete direct product (i.e. a a sector in modern terminology) is defined as an equivalence class between vectors belonging to the infinite direct product. This idea can be extended in the algebraic framework, and an essential feature is that the asymptotic information about the sector can be encoded in a concise way, without specifying the full algebra  (see e.g.  Chi-Keung Ng, ``On genuine infinite algebraic tensor products",  https://arxiv.org/abs/1112.3128). 

\bibitem{aqm} The litterature on algebraic QM is huge, see e.g. \cite{emch} for a classic textbook, and \cite{AQFT,OA,BF} for recent introductions and results; for more references see also \cite{csm4a,csm4b}. 

\bibitem{emch}  Gerard G. Emch, ``Algebraic Methods in Statistical Mechanics and Quantum Field Theory", 
Dover, New York, 2000 (reprint of the Wiley-Interscience 1972 edition).

\bibitem{AQFT}  C. J. Fewster and K. Rejzner, ``Algebraic Quantum Field Theory - an introduction", arXiv:1904.04051 [hep-th].

\bibitem{OA} Fernando Lledó, ``Operator algebras: an informal overview", arXiv:0901.0232 [math.OA]

\bibitem{BF} D. Buchholz and K. Fredenhagen, ``A $C^*$-algebraic approach to interacting quantum field theories",  arXiv:1902.06062 [math-ph].

\bibitem{balian}  A.E. Allahverdyana, R. Balian, T.M. Nieuwenhuizen, 
``Understanding quantum measurement from the solution of dynamical models",
Physics Reports 525:1, 1-166 (2013); arXiv:1107.2138 [quant-ph].

\bibitem{baez} J. Baez, ``Bell's Inequalities for C$^*$-Algebras", Lett. Math. Phys. 13, 135-136 (1987).

 \end{references}
\end{document}